# Pavlov's Dog Associative Learning Demonstrated on Synaptic-like Organic Transistors

O. Bichler[1], W. Zhao[1], F. Alibart[2], S. Pleutin[2], S. Lenfant[2], D. Vuillaume[2], C. Gamrat[1]

[1]CEA, LIST, Embedded Computing Laboratory, 91191 Gif-sur-Yvette Cedex, France.

[2]Institute for Electronics, Microelectronics and Nanotechnology (IEMN), CNRS, University of Lille, BP60069, Avenue Poincaré, 59652, Villeneuve d'Ascq, France.

**Abstract**

In this paper, we present an original demonstration of an associative learning neural network, inspired by the famous Pavlov's dogs experiment. A single Nano-particle Organic Memory Field Effect Transistor (NOMFET) is used to implement each synapse. We show how the physical properties of this dynamic memristive device can be used to perform low power write operations for the learning and implement short-term association using temporal coding and Spike-Timing-Dependent Plasticity (STDP)-based learning. An electronic circuit was build to validate the proposed learning scheme with packaged devices, with good reproducibility despite the complex synaptic-like dynamic of the NOMFET in pulse regime.

# 1 Introduction

In classical conditioning, associative learning involves repeatedly pairing an unconditioned stimulus, which always triggers a reflexive response, with a neutral stimulus, which normally triggers no response. After conditioning, a response can be triggered for both the unconditioned stimulus and the neutral stimulus, the later one becoming a conditioned stimulus. This concept goes back to Pavlov's experiments in the early 1900s. In his famous experiments, he showed how a neutral stimulus - like the ring of a bell - could be associated to the sight of food and trigger the salivation of his dogs (Pavlov, 1927). Associative memory is now a key concept in the learning and adaptability processes of the brain. If this ability is ubiquitous in everyday life, it is because it is likely a direct consequence of the structure and plasticity of the brain, down to the synaptic plasticity level. This is the basis of the Hebbian theory, which can be summarized as *cells that fire together, wire together*. It is therefore not surprising if associative learning is extensively studied in artificial neural networks (Hassoun, 1993; Norman, 2007).

However, the lack of an efficient implementation of artificial synapses for associative learning neural networks has greatly impeded the use of associative memory as a general-purpose type of memory or learning tool. The need of several transistors to implement dynamical synapses (Hynna and Boahen, 2006) is indeed not efficient enough compared to re-programmable digital logic, considering that at least several thousands of synapses are generally needed to perform any large scale information processing functionality such as associative memory on "real-life" data - the processing of visual or auditory stimuli for example (Bichler et al., 2012) - with limited reuse for other tasks. Synapses implemented with the most current CMOS technology would therefore still be several orders of magnitude behind their biological counterparts in terms of area and power consumption, which can be roughly estimated to ~ $10^{12}$ synapses/cm$^3$ and 1-10 fJ/spike if one considers an average firing rate of 1-10 Hz, given a power consumption of the human brain on the order of 10 W (Kandel et al., 2000). This is the reason why many new nano-devices currently actively researched to replace flash memory are receiving considerable attention from the neuromorphic engineering community.

Among them, Phase-Change Memory (PCM) devices have been proposed to realize artificial synapses implementing Spike-Timing-Dependent Plasticity (STDP), by using gradual crystallization and amorphization of the phase-change material, corresponding to gradual increase and decrease of conductance, for Long-Term Potentiation (LTP) and Long-Term Depression (LTD), respectively (Kuzum et al., 2012). In Conductive-Bridging RAM (CBRAM), the conductance of the device can be modulated by controlling the growth of the conductive filament in the material (Kund et al., 2005). If PCM crystallization has been show to be a truly cumulative process, meaning that successive identical programming pulses lead to gradual increase of conductance (Suri et al., 2012), it is not clear how this can be achieved with PCM amorphization and CBRAM, which require more complex programming schemes for synaptic applications (Bichler et al., 2012; Yu and Wong, 2010). While these two technologies are currently reaching industrial stage, multi-device synaptic applications are yet to be demonstrated in practice. Resistive RAM (RRAM or OXRAM), like Univ. Michigan nanoscale synapses (Jo et al., 2010) or HP $TiO_2$ memristors (Strukov et al., 2008), also present interesting characteristics for synaptic applications, with a cumulative effect for the conductance programming (Querlioz et al., 2011), as some carbon nanotube devices (Agnus et al., 2010; Zhao et al., 2010). All these devices fall in the class of memristive devices. Their high scalability potential and their programming scheme make them serious candidates to implement efficient artificial synapses.

Beside the above-mentioned technologies, volatile and/or organic synaptic memory devices have received comparatively little attention from the community. Bi-stable organic memories based on PMMA:ZnO (Ramana et al., 2012) or PMMA:C60 (Frolet et al., 2012) nanocomposites have been recently shown. Non-volatile multilevel conductance has also been demonstrated, for example in organic/Si nanowire transistor, by controlling the concentration of ions in the thin film layer through the gate voltage (Lai et al., 2008). Although the physical processes modulating the conductance might well be cumulative, the practicability of synaptic LTP or LTD implementation by gradual conductance change remains to be assessed. This leads us to the Nano-particle Organic Memory Field Effect Transistor (NOMFET) (Alibart et al., 2010; Bichler et al., 2010). It was shown in previous work (Alibart et al., 2012) that the NOMFET could be seen as a memristive device, by modulating the conductivity of its channel through the charging of nano-particles embedded into the organic semi-conducting channel. It is volatile and has a retention time of typically 10 to 1000 s. With these physical properties, the NOMFET can exhibit many behaviors of a dynamic synapse (Abbott et al., 1997; Tsodyks et al., 1998), when used in pulse regime, with a clear cumulative effect.

An elementary associative memory with memristive devices was first proposed by Pershin and Di Ventra, effectively reproducing the Pavlov's dogs experiment (Pershin and Di Ventra, 2010). The memristive devices were emulated with a micro-controller however, thus mitigating many difficulties that could arise from the interfacing with physical nano-devices, which are hard to predict in simulation with behavioral models. Another approach was proposed in (Ziegler et al., 2012) to realize the same elementary associative memory using a single Pt/Ge$_{0.3}$Se$_{0.7}$/SiO$_2$/Cu memristive device. In this paper, we propose an original scheme using NOMFETs to implement dynamic associative learning and demonstrate it by interfacing NOMFETs to a CMOS discrete circuit, thoroughly described in the "Methodology" section. We show how the unique synaptic properties of the NOMFET can be used directly at the device level to implement what we call a dynamic associative memory, where the association is only retained as long as there is a minimal activity at its input. In the final "Experiments and Results" section, the learning dynamic is viewed in relation to the NOMFET volatility, the learning scheme is interpreted in terms of STDP and the impact of variability is briefly discussed.

## 2 The NOMFET

The NOMFET is a three-terminal device, as the conventional MOSFET (see figure 1). As an organic transistor, it features the classical p-type transistor behaviors in accumulation regime, but with added memristivity. For a fixed gate voltage, its conductivity can indeed be modulated by charging or discharging the gold nano-particles (NPs) embedded in its channel, with a negative or positive gate voltage respectively. When a negative gate voltage is applied, the NPs are positively charged and the repulsive electrostatic interaction between the holes trapped in the NPs and the ones in the pentacene p-type channel therefore reduces the channel conductivity. The charge retention time is typically of 10 to 1000 s. The programming of the conductivity of the channel can be done with no current flowing from the source to the drain, by keeping $V_{DS} = 0$.

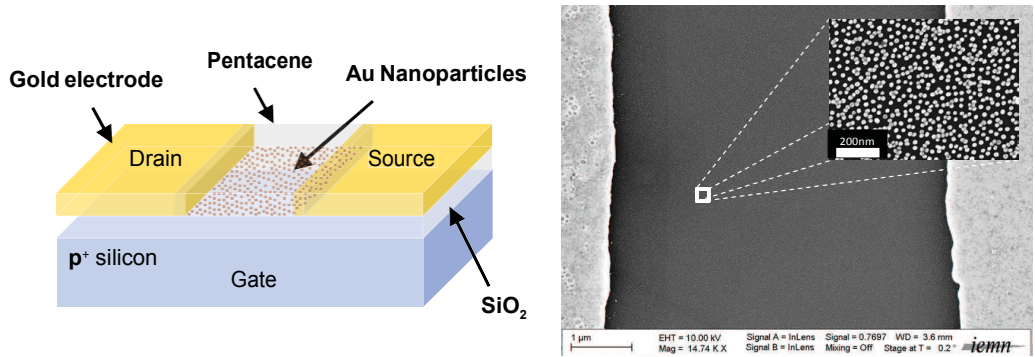

*Figure 1*: *Physical structure of the NOMFET transistor (left). It is composed of a $p^+$ doped bottom-gate covered with silicon oxide (200 nm thick). Source and drain electrodes are made of gold and Au NPs (20 nm in diameter) are deposed on the inter-electrode gap (5 μm), before the pentacene deposition. Scanning electron microscope image of the NP array between the source and drain electrodes (right).*

Figure 2 shows the charging/discharging dynamic characteristic of the NOMFET. The current is measured before and after a 10 s programming pulse is applied on $V_G$. The $\Delta I/I - V$ curve shows the change in conductivity in function of the pulse voltage, with a non-linearity between 0 and 15 V. This characteristic remains the same when no current flows through the channel, thus making the NOMFET an ideal memristive device (Alibart et al., 2012) in the sense that on average, no power is dissipated for the programming of its conductivity. The evolution of the conductivity is essentially controlled by the time integral of the gate voltage, contrary to the memistor (not to be confused with the memristor), which was programmed by the time integral of the current flowing through its third terminal (Widrow et al., 1961).

While we have reported synaptic-like behavior in short (200 nm gap, 5 nm NPs) NOMFET working at a bias of -3 V (Alibart et al., 2010), here we use, for the sake of demonstration, larger (5 μm gap, 20 nm NPs) NOMFET because they showed the largest plasticity amplitude (i.e. the largest modulation of the NOMFET conductivity under the application of programming pulses). With an ON-state current ranging from 50 to 500 nA and a drain-to-source voltage of 15 V, the equivalent resistance of the synapse ranges from 30 MΩ to 300 MΩ. While this is relatively high compared to other memristive technologies (Bichler et al., 2012; Kund et al., 2005), this is accomplished without sacrificing the OFF/ON resistance ratio thanks to the transistor field effect (OFF-state current below 0.1 nA, measurement limited). This makes the NOMFET low power, yet still usable in device arrays for large-scale networks (Liang and Wong, 2010).

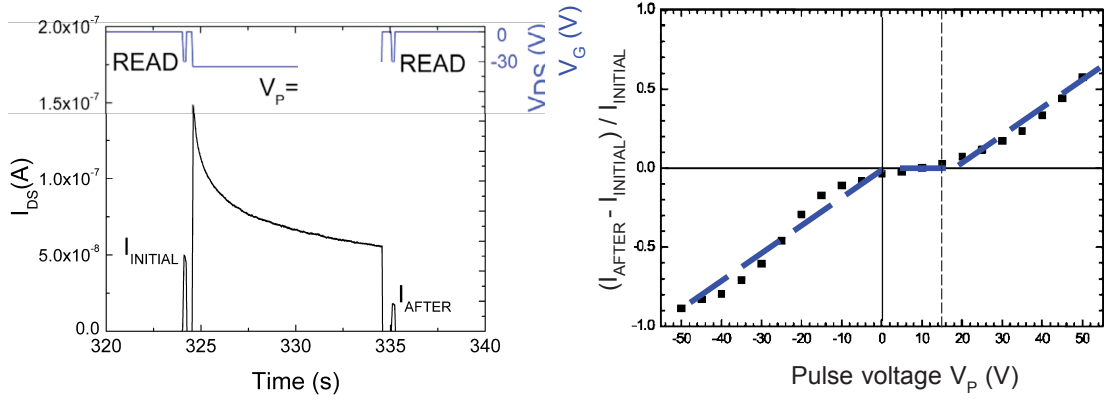

*Figure 2*: Dynamic characteristic: relative conductivity change as a function of the pulse voltage (dashed line is a guide for eyes). The current is measured just before and after the application of a 10 s pulse with different amplitudes. The relative change in the current is related to the amount of charges in the NPs. For negative pulses, the NOMFET is in accumulation, NPs are positively charged by holes, thus reducing the current by coulombic effect. For positive bias, holes are detrapped, leading to an increase in the current. The curve is non-linear with a zone between 0 V and 15 V where charges in the NPs are not affected by the applied bias. For more details, see ref. (Alibart et al., 2012).

## 3  Methodology

Our associative memory is constituted of two input neurons, two synapses and one output neuron (figure 3). A synapse is implemented using a single NOMFET and the neurons are built with discrete CMOS devices on a custom electronic board. The association is realized relying solely on the plasticity behavior of the NOMFET: its conductivity is changed in an unsupervised way by the interaction of the pre-synaptic pulses (input signal) and the post-synaptic pulses (feedback), such that the association is made when there is a temporal correlation between the pre- and post-synaptic pulses for a long time (see how this relates to STDP in the "Experiements and Results" part).

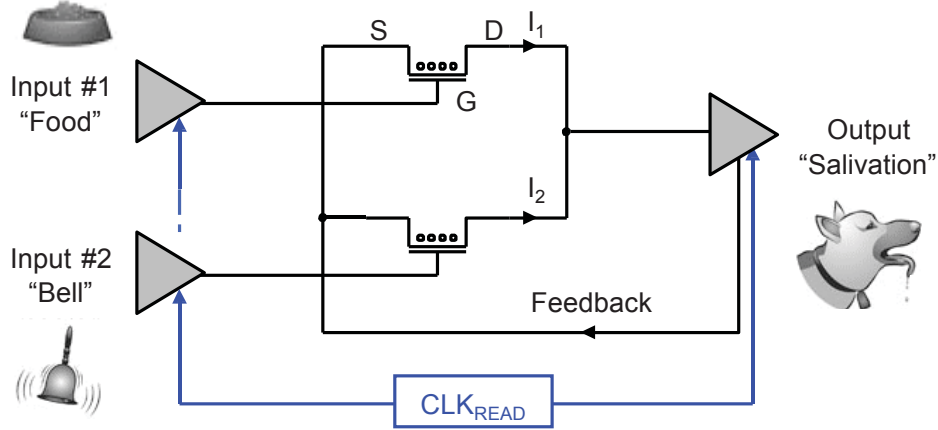

*Figure 3*: *Equivalent electronic circuit for the associative memory. There are 3 neurons (Input #1, Input #2 and Output) and 2 synapses/NOMFETs. A global clock signal (CLK$_{READ}$) enables synchronization of the input and output neurons during the read operations. I$_1$ and I$_2$ are the measured drain-source currents.*

The operations on the synapses are done in two steps, a "read" step and a "write" step. Separating these two steps allows a greater control on the dynamic of the system, effectively decoupling the physical plasticity of the NOMFETs from the associative learning dynamic. More generally, even in non-volatile memristive devices based neural networks, separating these two steps can have several advantages: because the read can be done with a lower voltage and/or a shorter pulse, the conductivity change of the device induced by the read can be minimized, as does the power consumption. Another interest of this approach is that current does not need to flow through the device during the write step. Indeed, contrary to resistive-based memristive devices, the nano-particles in the NOMFET channel can be charged or discharged without the transistor being in the ON state, thus modulating its conductivity with no current flowing from the source to the drain. We effectively use this principle in our experimental setup, where the pre-synaptic pulses are applied to the gate of the NOMFETs. The output neuron is connected to the source and drain terminals of all the devices and can apply separate voltages to the source and to the drain.

## 3.1 Read and Output Neuron Activation

A read step occurs periodically and triggers the activation of the output neuron if an input is present on a highly conductive synapse (meaning that the neuron is currently "listening" to that synapse). During a read step, the conductive state of the devices is measured. A feedback is generated if there is an active input signal at one of the NOMFETs during the read and that it has a conductivity higher than a fixed threshold. The read step implies that both input and output neurons are synchronized, as the active input neurons should turn on the NOMFETs by applying a negative voltage to the gate. The output neuron should apply a negative voltage pulse to the source of the devices and measure the drain-source current at the drain terminal. This process can be done sequentially by turning on only one NOMFET at a time so that all the drain and all the source terminals can be linked together.

## 3.2 Write (Input and Feedback Interaction)

The conductivity of the NOMFETs is modulated during the write steps, through the interaction of the pre-synaptic pulses (gate) and the post-synaptic pulses (source and drain). In order to achieve an associative memory, the conductivity of a NOMFET should increase only in the occurrence of simultaneous pre- and post-synaptic pulses, meaning that there is a correlation between the events leading to the activation of the output neuron and the input events coming to the synapse. This is implemented by ensuring that only the interaction of pre- and post-synaptic pulses leads to a significant increase in the conductivity of the NOMFET, by applying a higher voltage across the gate and source-drain terminals of the device than with a pre- or a post-synaptic pulse alone. Since the relation between the conductivity change and the applied voltage amplitude is non-linear (see figure 2), it is possible to effectively maximize the effect of the two interacting pulses and minimize the effect of a single pulse to the conductivity. The shape of the pulses is given in figure 4:

- When no feedback is present, the input only has limited effect on the conductivity of the synapse: no input means a constant voltage of -15 V is applied to the gate, which tends to decrease the conductivity over time (loss of memory). This is mitigated by the time when the input is being active, where the conductivity slightly increases. The overall effect is that the conductivity of the synapse is maintained as long as there is a sustained activity at its input.

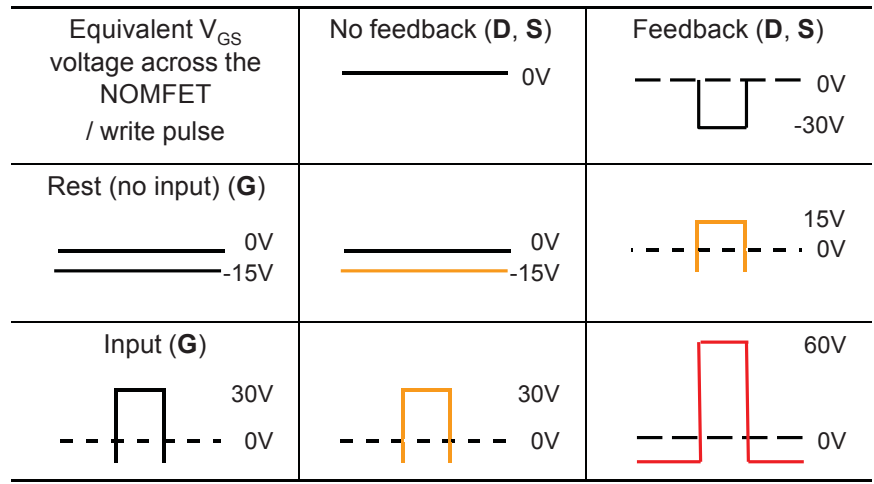

*Figure 4: Input and feedback pulse shapes to implement the associative learning. Input signals are applied to the gate and feedback to the source and drain terminals. The inputs are negatively biased, meaning that a -15 V voltage is constantly applied to the inputs when no signal is present. Only the interaction of an input signal and a feedback signal creates a programming voltage between the gate and the drain-source terminals large enough to significantly increase the conductivity of the NOMFET, thus leading to the association when several interactions occurs in a short time.*

- When a feedback is present, there is no change in conductivity in the absence of input, whereas a major increase in conductivity occurs upon the interaction of the input present pulse and the feedback pulse. The -15 V bias was chosen in order to avoid maintaining an association with only the feedback (thus the voltage across the NOMFET is 15 V with the feedback, instead of 30 V with an input).

During write operations, the source and drain terminals are connected together, which implies that the only electric potential difference in the device is across the gate, regardless of the simultaneity of the input and feedback pulses. In this configuration, the device act as a capacitor, with the charging and discharging of the embedded NPs in the channel modulating the conductivity of the transistor, for a given gate voltage during read operations. Like a capacitor, because no current can flow through the gate, the change in conductance does not dissipate any power on average. This is unlike non charge-based, non-volatile, resistive memory devices like PCM or CBRAM, where conductance change requires Joule heating to induce phase-change or ions migration in the material. In these technologies, the programming current typically falls in the mA or $\mu$A range (Bichler et al., 2012; Kund et al., 2005).

## 3.3 Calibration

In a dynamical associative learning network, a calibration step is necessary to initialize the state of the synapses, in order to enable system responsiveness for an initial set of unconditioned stimuli. An optimal threshold can also be automatically computed for this initial configuration. Because the synaptic weights are programmed in parallel, this initial set-up does not depend on the number of synapses.

In our demonstration, this calibration process is automated in the micro-controller implementing the neurons (see next sections). It sets up the threshold before the associative memory learning is run. In addition, a constant gain is automatically applied to the reading of one of the two NOMFETs in order to compensate for the mismatch between the two devices. The calibration sequence is the following: (1) Measure the conductivity state of each device for the four possible programming states (with or without pre-synaptic spike and with or without post-synaptic spike). (2) Compute the gain to be applied to the second device to compensate for the mismatch. (3) Compute the optimal threshold. (4) Initialize the conductivity state of both devices.

Although steps (1)-(2) may be unpractical in larger systems, mismatch compensation for the static $I_{DS}$ characteristic may be eliminated for matched devices on the same die fabricated in controlled industrial process. Indeed, we found that most of the mismatch is introduced by the wire bounding during device packaging.

## 3.4 Experimental board

A custom electronic board was build to interface with the NOMFETs. The NOMFETs were packaged in a standard through-hole TO (Transistor Outline) metal can package. The design of electronic boards with off the shelf components to interface with experimental devices often proves to be challenging, not to mention specific constrains on temperature or atmosphere control for the wire bounding and packaging depending on the materials used. For the NOMFET, one has to be capable of measuring current in the nA range with voltage as high as ±30 V for the first generation of NOMFETs that we used for this setup.

Despite these possible difficulties, the presented NOMFETs are working in atmospheric condition without any special encapsulation and have shown a stability in time over more than 6 months. And we have previously reported that the synaptic properties of the NOMFET were maintained by decreasing the gate thickness down to 10 nm, reducing the gap size down to 200 nm and the NP diameter to 5 nm (Alibart et al., 2010), thus reducing the maximum programming voltage down to 3 V. The bottom up approach and the low temperature deposition condition for the fabrication of the NOMFETs is compatible with CMOS and we think that interfacing of experimental devices with standard electronic is a valuable step towards larger scale prototypes and co-integration with CMOS.

As the proposed associative learning scheme only requires four different voltages to generate the read and write pulses, the pulses at each terminal are generated with an analog multiplexer, the MAX14752, which can switch voltages up to ±36 V. Because we use the dynamic of the NOMFET in the ms-s range, the switching speed is not critical.

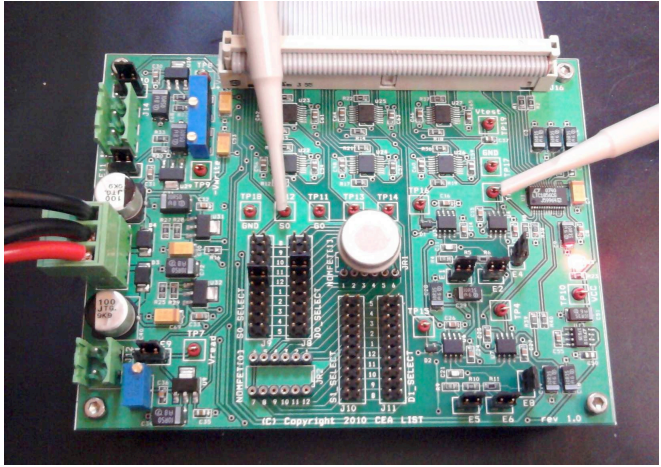

*Figure 5*: *Photography of the experimental board, with the NOMFETs in a TO (Transistor Outline) case at the center. The computer interface, including the FPGA (Field-Programmable Gate Array)-based control board implementing the control logic for the neurons, are not shown.*

For the read step, the conductivity is obtained by measuring the current flowing through the drain terminal when a positive voltage pulse is applied to the source, which is equivalent to applying a negative pulse to the gate and the drain. The current measurement is done with a current-to-voltage converter in a feedback-ammeter topology. In this topology, the operational amplifier delivers a voltage proportional to the current flowing through the feedback resistor, which is equal to the drain-source current of the NOMFET. It is particularly adapted for the measurement of small currents (below the $\mu$A). The OPA445 is used for this task, which is one of the few operational amplifiers to support a large supply voltage of up to ±45 V and to have a low input bias current (of the order of 10 pA), required to measure current down to the nA. It is followed by a differential amplifier AD629B to allow measurement of the current when a feedback pulse is applied to the NOMFET, by suppressing the common mode voltage of the feedback pulse. It supports a high common mode voltage and has a low input voltage offset. Finally, the LTC1856 analog-to-digital converter (ADC) is used to obtain a digital value of the current. It can measure bipolar voltages up to ±10 V directly applied at its input and can withstand input voltages as high as ±30 V without damage. It also has eight multiplexed measurement channels with a resolution of 16

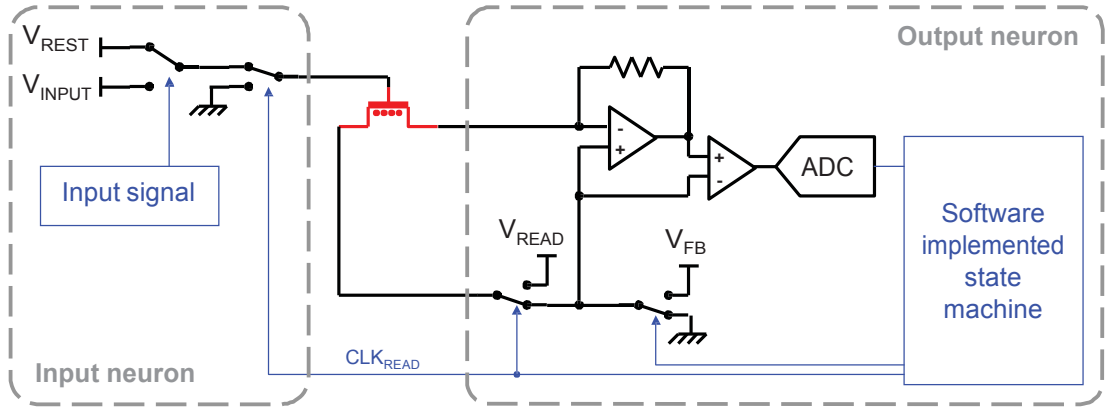

*Figure 6*: *Simplified schematic of the experimental setup for one NOMFET. The blue/thin parts are implemented in software on the Microblaze synthesized on the FPGA. In read mode, the source terminal is connected to $V_{READ}$ while the other two terminals are grounded. When the output neuron activates, the source and drain terminals are connected to $V_{FB}$ and the device's conductivity is modulated depending on the gate voltage $V_{REST}$ (no input) or $V_{INPUT}$ (input).*

bits, making it easy to measure several devices with a single component.

A photography of the board is shown in figure 5 and a simplified schematic of the circuit is shown in figure 6. No special care was needed to measure current down to the nA, other than careful layout with proper ground plane and decoupling capacitors.

## 3.5 Neuron emulation

While the experimental board provides the digital-to-analog (pulses generation) and analog-to-digital (current measurement) interface to the NOMFETs, the neurons themselves are implemented in digital logic on a FPGA (Field-Programmable Gate Array)-based board, which does the initial calibration and supervises the learning process. Instead of implementing the control logic directly as a state machine in a hardware description language, a programmatic approach was preferred for flexibility. A Xilinx Microblaze (Xilinx, 2012) micro-controller core was therefore synthesized on the FPGA and programmed in the C language.

# 4 Experiments and results

To demonstrate the working order of our associative memory, we reproduced the Palvov's dog experiment. Figure 7 shows the association process: the two inputs of the memory models the sight of food and the hearing of a bell respectively, whereas the output neuron activity models the salivation of the dog. At first, the conductivity of the devices was programmed so that only the sight of food (unconditioned stimulus) triggers the activation of the output neuron (salivation). Before the association is made, the hearing of the bell (neutral stimulus) does not trigger the salivation, as the conductivity of the corresponding synapse is below the threshold of the output neuron. When both inputs are active simultaneously, the conductivity of the second synapse increases until it reaches the threshold (conditioning): the association is made. From this point, the hearing of the bell alone triggers the activation of output neuron (conditioned stimulus). In figure 8 we can verify that without feedback from the output neuron, the conductivity change implied by the pre-synaptic pulses alone is not enough to create the association.

Our associative learning circuit is symmetrical, which means that there is no difference between the input receiving the conditioning stimulus (sight of food) and the input receiving the neutral stimulus (bell sound). The conditioning input is the one that is initially programmed to have a synaptic weight above the output neuron threshold. This is in contrast with the asymmetrical scheme introduced in (Ziegler et al., 2012), where the two inputs cannot be interchanged.

## 4.1 NOMFET Volatility and Learning Dynamic

Because the conductivity of the NOMFET is volatile, what we really implemented is a short-term associative memory. Indeed, when no sustained activity is present at the inputs, the association is lost (figure 9). This functionality is directly provided by the physics of the device. There is no need for additional circuitry that would periodically decrease the conductivity of the synapses. We argue that this functionality is essential for any practical use of associative memory, because otherwise, the synaptic weights in such a system would eventually saturate due to the random short-time correlations between inputs.

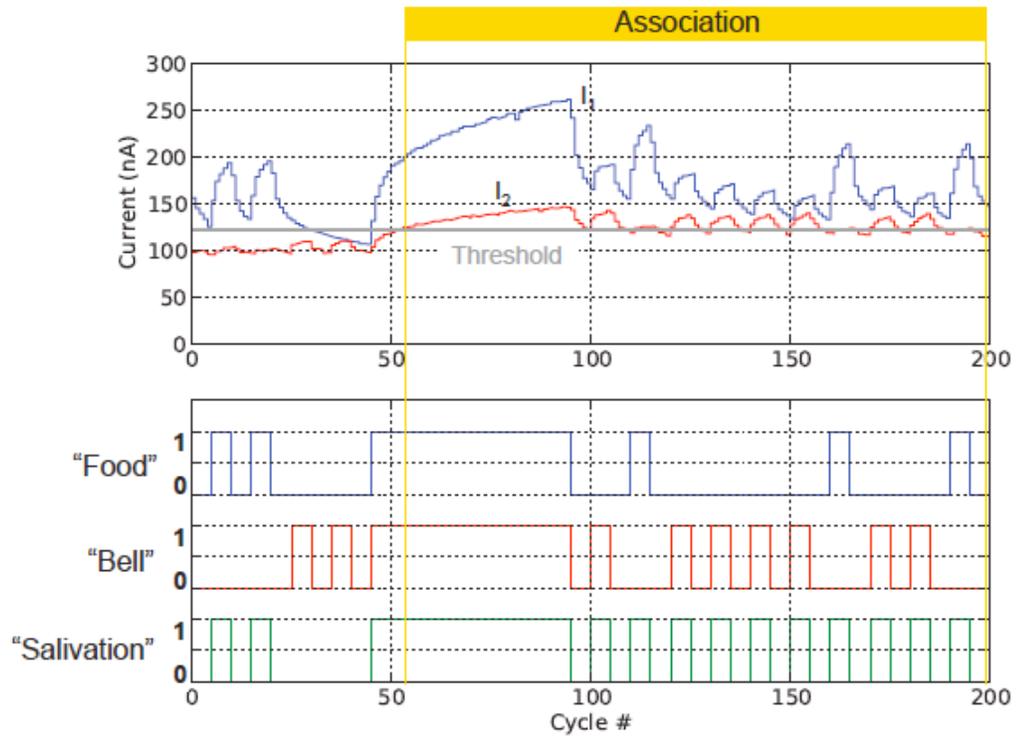

*Figure 7*: *Dynamic associative learning: association is maintained as long as there is a minimum activity at the inputs. The duration of a cycle is 200 ms. After each cycle, the current is measured and compared to the neuron's threshold. In this experiment, the minimal duration of an input pulse is 5 cycles (1 s). The total duration of the sequence is 40 s.*

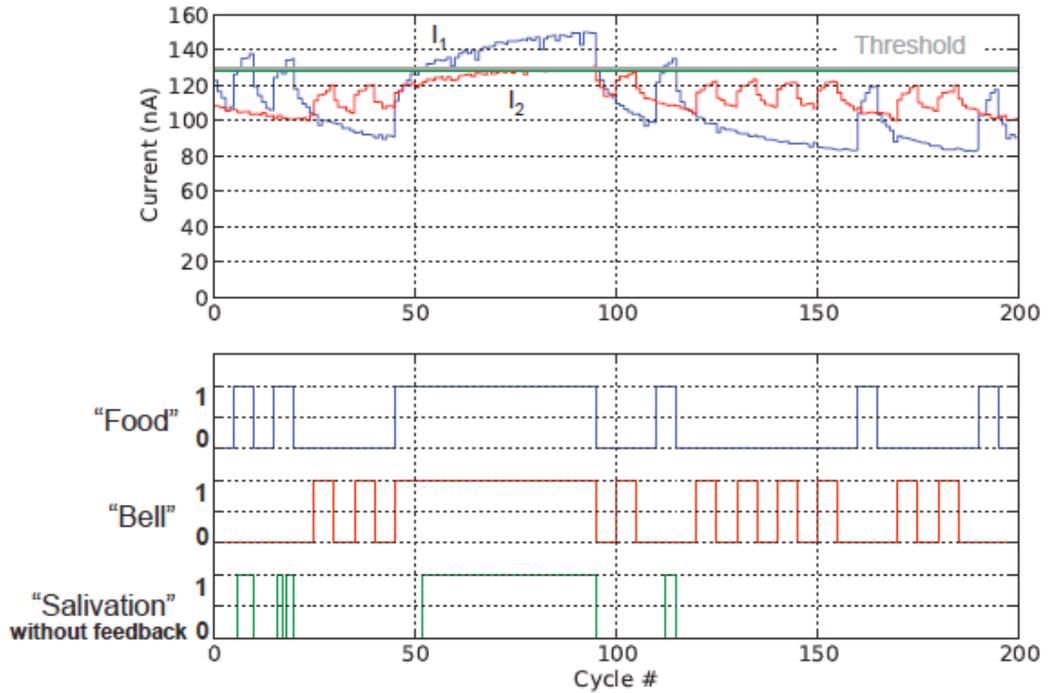

*Figure 8*: *Same sequence, with disabled feedback: no association is made and the first synapse loses the association as well. As in the other experiments, the best threshold is automatically computed at the beginning of the sequence using the same calibration procedure, in which the feedback was also disabled.*

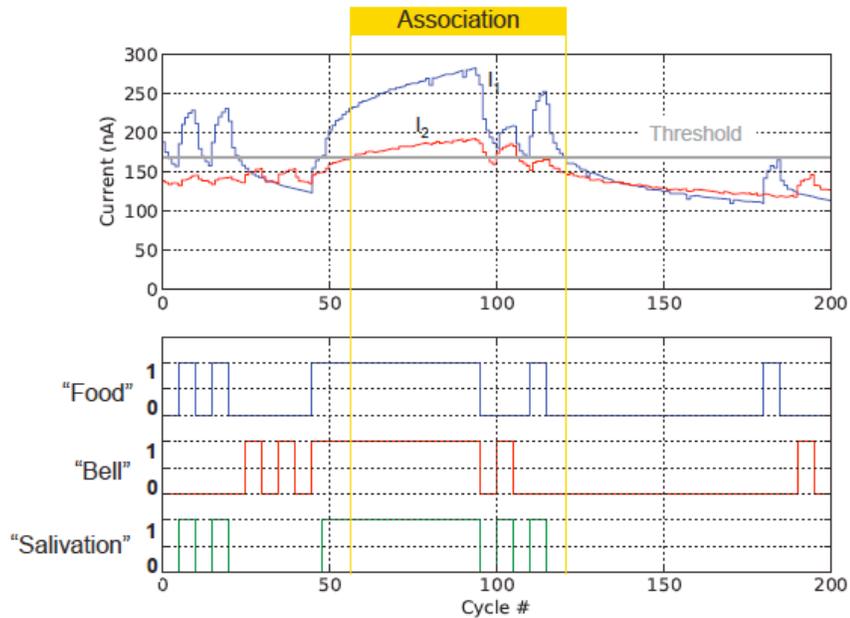

*Figure 9*: *Dynamic associative learning: if no sustained activity is present at the inputs, the association is lost.*

The NOMFET exhibits both a short-term and a long-term dynamic, that are clearly identifiable in figure 7. The short-term dynamic corresponds to the fast increase and decrease of conductance upon activation and deactivation, respectively, of the input neuron (typical time dynamic ~ 1 s). The average conductance or current evolution during the learning follows a longer term dynamic (typical time dynamic ~ 10 to 100 s): the average current is indeed higher after the association than before. A long potentiating pulse (conditioning step) is required to induce this more durable change. Such dynamical synapse in time is likely to be impractical to emulate with non-volatile memristive devices, like PCM, as each device would require some external time reference to implement its time dynamic.

## 4.2 Relationship with STDP

Spike-Timing-Dependent Plasticity is currently one of the most widely studied neuromorphic learning algorithm on memristive devices (Yu and Wong, 2010; Kuzum et al., 2012; Jo et al., 2010; Bichler et al., 2012). There is not a single STDP rule however: a broad family of synaptic update characteristics in function of the pre-post synaptic time difference were recorded (Wittenberg and Wang, 2006). Most notably, asymmetric STDP was observed in hippocampal glutamatergic synapses (Bi and Poo, 1998) and symmetric STDP was observed in CA1 region of the hippocampus (Nishiyama et al., 2000) and GABAergic synapses (Woodin et al., 2003).

Asymmetric STDP with NOMFET was reported in (Alibart et al., 2012). The programming scheme realized in our associative memory however emulates a form of symmetric STDP learning rule, as shown qualitatively in figure 10. Indeed, a significant increase in conductivity is only induced when there is a simultaneous occurrence of pre- and post-synaptic events (input and feedback pulses), leading to a 60 V equivalent pulse across the NOMFET.

## 4.3 Impact of variability

Although there is a factor 10 in the mean conductivity between the two NOMFETs used in this experiment (before applying a correcting gain), the variability on the dynamical behavior of the devices is low in comparison (Bichler et al., 2010; Alibart et al., 2012). Because our associative memory is based on temporal coding, only the relative variation of the conductivity obtained through programming pulses and natural charge relaxation of the NPs imposes the dynamic of the learning. In this regard, STDP-based learning systems were shown to be tolerant to variations from 20% up to 100% in synaptic weight update steps (Querlioz et al., 2011; Bichler et al., 2012), which was confirmed by the high reproducibility of our experiment. Incidentally, the current variations for the two NOMFETs are not necessarily matched during the calibration process, which is apparent in figures 7-9. Instead, the effect of the calibration process is to match the temporal dynamic of the conductivity variations. This dynamic remains valid for longer time scale, with cycle duration up to one order of magnitude higher.

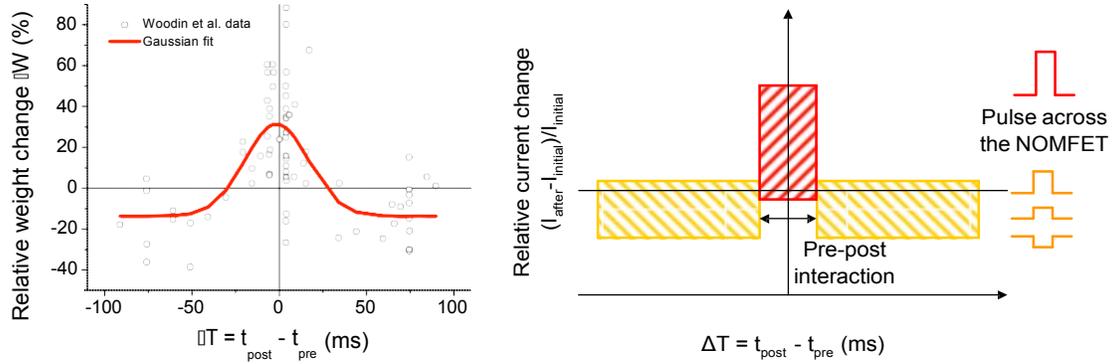

*Figure 10*: *Left: relative synaptic weight change in symmetric STDP, observed in GABAergic synapses. Experimental data from (Woodin et al., 2003). Right: relative conductance change realized in our associative learning for a NOMFET synapse, depending on the pre-post programming pulses interaction (qualitative diagram).*

## Discussion and Conclusion

Associative memory is a fundamental computing block in the brain and is implemented extremely efficiently in biological neural networks. An efficient and scalable implementation of associative memory would certainly benefit many applications, especially in the area of natural data processing. One could imagine, using the same principle as the Pavlov's dog learning, association of visual and auditory features, but directly below artificial retina and cochlea event-based sensors.

We demonstrated experimentally an elementary associative memory, which uses only one NOMFET memristive nano-device per synapse. Although it may not be able to ultimately achieve the scalability of biological synapses, it exhibits dynamical behaviors closer to biology than any other known memristive device. Furthermore, the volatility of the NOMFET that leads to the dynamic learning scheme introduced in this paper is not necessarily irrelevant when considering biological synchronous neural models for short-term memory (Ward, 2003).

## Acknowledgments

We thank the following colleagues for help in the device fabrication: K. Lmimouni (pentacene deposition), D. Guerin for surface chemistry (grafting of NPs), J. Dubois (earlier electronic circuit prototyping), and D. Querlioz for helpful discussions. Financial support: European Union through the FP7 Project NABAB (Contract FP7-216777).